\documentclass[letter,scriptaddress,twocolumn, prl]{revtex4}
\usepackage{amsmath,amssymb,latexsym}
\usepackage{wrapfig}
\usepackage{subfig}
\usepackage{times}
\usepackage{epsfig}
\usepackage{graphicx}
\usepackage{color}
\usepackage{dcolumn}
\usepackage{bm}
\usepackage{setspace}
\usepackage{enumitem}
\usepackage{mathtools}
\usepackage{caption}
\setlist[itemize]{noitemsep}

\newcommand \ofr {\left(\textbf{r}\right)}
\newcommand \ofri[1] {\left(\textbf{r}_#1\right)}
\newcommand \br {\textbf{r}}
\newcommand \be {\begin{equation}}
\newcommand \ee {\end{equation}}
\newcommand{\bk}{\textbf{k}}

\begin{document}

\title{New density functional approach for solid-liquid-vapor transitions in pure materials}

\author{Gabriel Kocher$^1$}
\author{Nikolas Provatas$^1$}

\affiliation{$^1$ Department of Physics, Centre for the Physics of Materials, McGill University, Montreal, QC, Canada\\}

\date{\today}

\begin{abstract}
A new phase field crystal (PFC) type theory is presented, which accounts for the full spectrum of solid-liquid-vapor phase transitions within the framework of a single density order parameter. Its equilibrium properties show the most quantitative features to date in PFC modelling of pure substances, and full consistency with thermodynamics in pressure-volume-temperature space is demonstrated. A method to control either the volume or the pressure of the system is also introduced. Non-equilibrium simulations show that 2 and 3-phase growth of solid, vapor and liquid can be achieved, while our formalism also allows for a full range of pressure-induced transformations. This model opens up a new window for the study of pressure driven interactions of condensed phases with vapor, an experimentally relevant paradigm previously missing from phase field crystal theories.
\end{abstract}

\maketitle

%
In the study of materials, modelling non-equilibrium phase transformations is crucial, and requires capturing atomic length features, while remaining consistent with Thermodynamics at long length and time scales. Toward this goal, phase field crystal (PFC) modelling \cite{elder2002} has recently emerged as an efficient and mathematically accessible option, incorporating the thermodynamics of phase transformations and most salient solid state properties, including elasto-plastic deformations and grain boundaries, all on diffusive timescales~\cite{emmerich2012}. Extensions to the original model have been applied to complex structural transformations in pure materials
~\cite{greenwood2010,mkhonta2013}, multi-component alloys~\cite{ofori-opoku2013} and the study of solid-liquid and solid-solid transformations~\cite{Mellenthin08,berry2014,berry2011}. 

To date, however, most PFC modelling has considered only liquid-solid or solid-solid transitions at fixed average density, a situation that severely precludes the applicability of the PFC paradigm to problems related to the interaction of condensed phases with vapor. A method to model such systems was introduced~\cite{voorhees2013}, but it is not derived from a single order parameter, and precludes a description of the critical point. In this letter we introduce a new, more fundamental PFC-type theory of pure substances, which accounts for the full spectrum of solid-liquid-vapor transitions within the framework of a single density order parameter.  Our formalism is shown to be fully consistent with thermodynamics in Pressure-Volume-Temperature space,  while inheriting the features of previous PFC models. It also naturally accounts for different anisotropies and nucleation barriers for vapor/solid and liquid/vapor systems. We additionally introduce a method to control either the volume or the pressure of the system. As a demonstration, we show an application in pressure-driven phase transformations.

Consider classical Density Functional Theory (c-DFT)~\cite{ram1979,archer2006}: Let $\rho\ofr$ be a field representing the atomic density of an interacting liquid. The free energy of such a liquid is generally written as $F_{cdft}\left[\rho\right]/(k_BT)=F_{id}\left[\rho\right]+\Phi \left[ \rho\right]$ where $F_{id}$ is the energy of an ideal gas and $\Phi$ the contribution due to interactions. $\Phi$ is then treated by functional expansion around a reference density $\bar\rho$, in a power series of $n=(\rho-\bar{\rho})/\bar{\rho}$, and interactions are described by a sequence of n-point correlations $C^{(n)}(\br_1,..,\br_n)$. 
%
%
%
While these correlation functions are not known in general, a truncation of the series to second order along with a suitable ansatz of $C^{(2)}$ has been shown to separately describe both vapor-liquid interfaces~\cite{evans1979} or solidification problems~\cite{ram1979} with success. PFC methods additionally rely on an expansion of the ideal free energy around $n\sim0$, to create what one may call a "smooth atom" approximation~\cite{elder2007,jin2006,Granasy10,Wu10a} of an atomic density field. While the atomic density interpretation is lost, the order parameter field $n$ still exhibits spatial variations and retains numerous crucial features of the c-DFT atomic density.

To overcome the limitations of two-point correlations on multi-phase behaviour, we introduce here a theory that relies on higher order correlations. Consider the Van der Waals theory for the liquid vapor transition~\cite{plischkebergersen}. Its improvement to the Ideal gas law is based on two simple mean-field postulates: the attraction between particles is proportional to the average surrounding density, and each particle proportionally reduces the free volume available to other particles. At the field theory level for the spacially varying coarse grained field $\rho$, such improvements can be described by the free energy 
$F_{VdW}[\rho] /(k_BT)=F_{id}-\int \text{d}\br\left[\rho_{mf}\text{ln}(1-\rho_{mf} b)+\frac{a}{k_BT}\rho_{mf}^2\right]$,
where $\rho_{mf}\ofr=\int\text{d}\br\chi(\br-\br')\rho\ofr$ is a local spacial average of the density field $\rho$, with $\chi$ a local smoothing kernel. In the limit of a fully uniform field and setting $\rho=\rho_{mf}=N/V$, this free energy reduces to the standard Van der Waals free energy, where $a$ and $b$ respectively control the magnitude of the attraction and repulsion between atoms. This formulation lends itself to an interesting c-DFT interpretation. Indeed, expanding it around a reference density generates a power series in $\rho_{mf}$, that we may interpret as a series of correlation functions. In the following we introduce a formulation that incorporates all the qualitative contributions from the Van der Waals theory into the free energy of the standard PFC-expanded  formalism. In addition of an expanded ideal free energy, it contains both a sharp 2-point kernel and a set of long-range kernels, which allow for the description of solid, liquid and vapor phases from a single microscopic order parameter field.

Our model uses the following free energy functional ($\mathcal{F}=F/\bar{\rho}k_BT$):
\begin{eqnarray}
\mathcal{F}\left[n\right]&=&\int \text{d}\br\left[ \frac {n\ofr^2} 2 -\frac {n\ofr^3} 6+\frac {n\ofr^4} {12}\right]\nonumber\\
&-&\!\!\! \frac 1 2 \int \text{d}\br_1\text{d}\br_2 ~ C^{(2)}(\br_1-\br_2)n\ofri{1}n\ofri{2} \label{free}\\
&+&\!\!\!\sum_{m=3}^4 \frac {1} {m} \left(\int \text{d}\br_1..\text{d}\br_m  \chi^{(m)}(\br_1,..,\br_m)n\ofri{1}..n \ofri{m} \right)\nonumber
\end{eqnarray}
The first line results from the expansion of the ideal gas free energy $F_{id}\left[\rho\right]$, while the second line adds a multi-peaked 2-point correlation function. The choice of the latter term determines the structure and properties of the solid phase. While elaborate choices can be made for this term (to target specific 2D or 3D structures~\cite{greenwood2010,mkhonta2013}), for simplicity we choose a kernel that yields triangular/BCC structures in 2D/3D~\cite{elder2002}:
$
C^{(2)}(\br_1-\br_2)=1-r-B_x(1-\nabla^2)^2
$.
\begin{figure}
\centering
\includegraphics[width=0.5\textwidth]{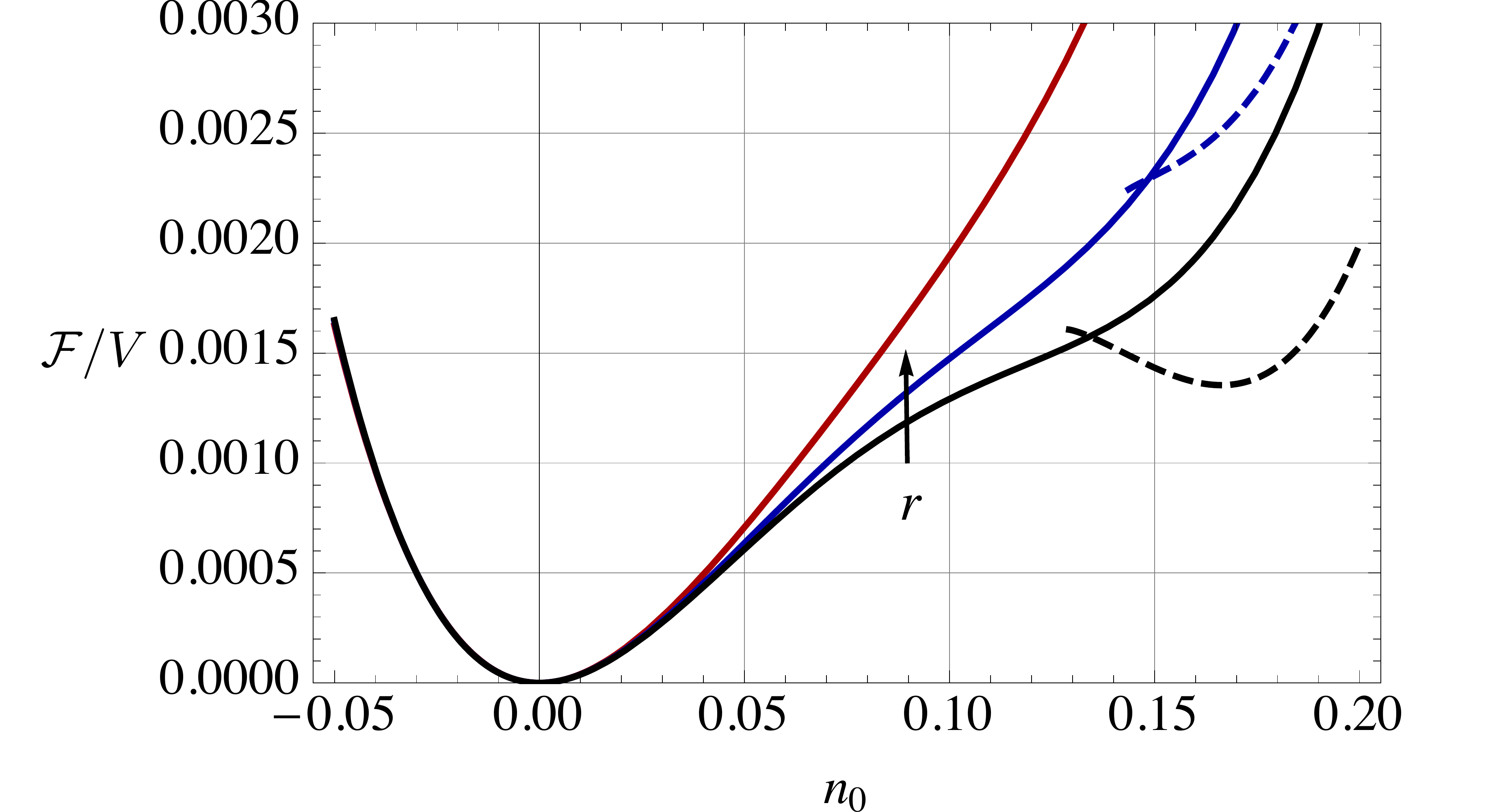}
\caption{2D Free energy landscapes (vs. average density) for different effective temperatures $r$. Uniform phases: continuous lines, periodic phases: dashed. Black: $r=0.14$, blue: $r\approx0.148$ (triple point), red: $r=0.17$. Other parameters: $a = 50$, $b = -19$, $c= 50$, $B_x=0.7$. \label{Wells}}
\end{figure}
\begin{figure}
\centering
\vspace{-15pt}
\includegraphics[width=0.5\textwidth]{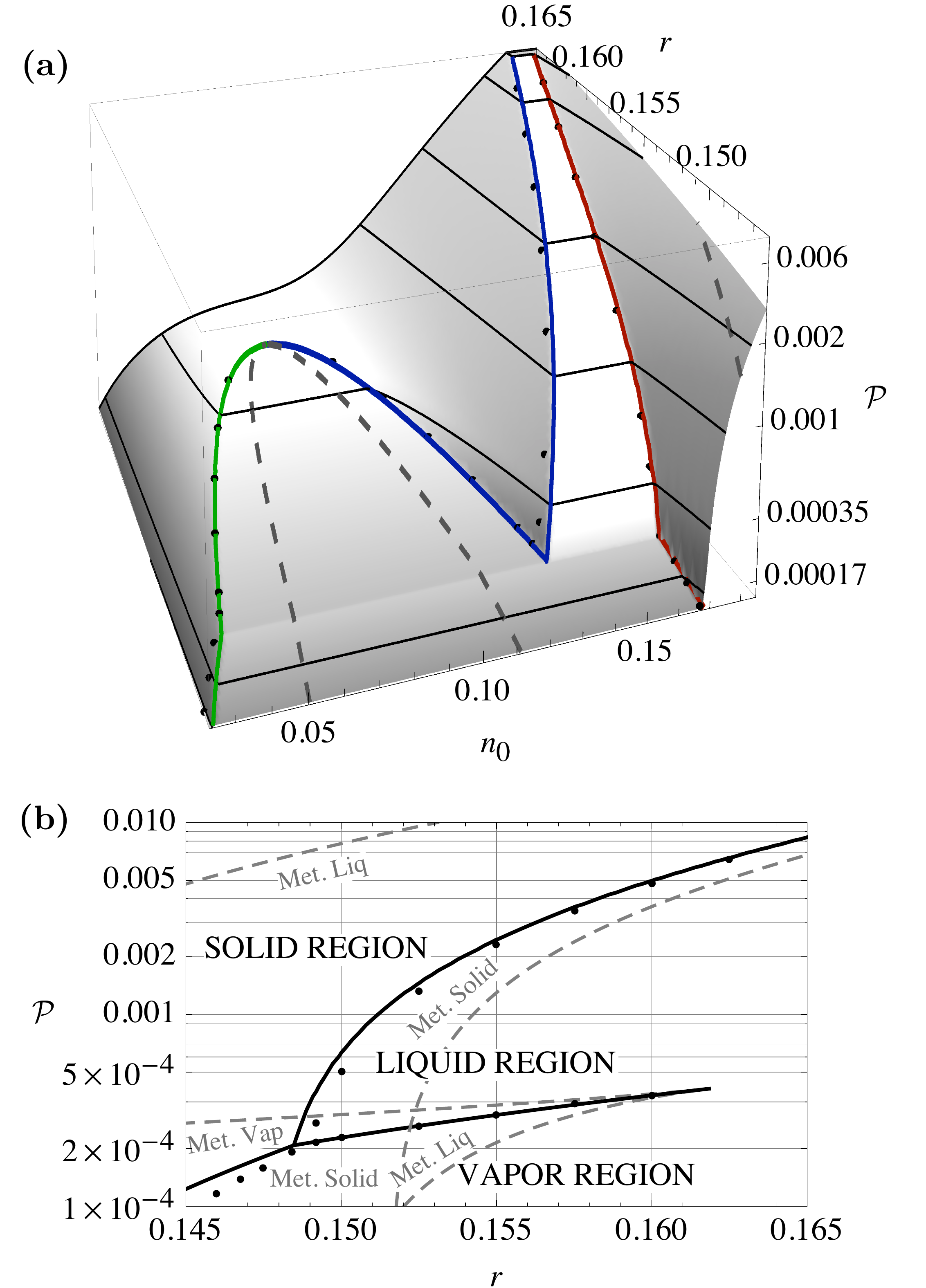}
\caption{(a) Density-temperature-pressure and (b) pressure-temperature phase diagrams of model. Solid thick lines correspond to 1-mode calculations. In (a), green: vapor, blue: Liquid, red: periodic coexistence values. Dashed line is the vapour-liquid spinodal. In (b), dashed lines show metastability regions. Black dots show average coexistence density (in (a)) or pressure (in (b)) from isothermal simulations. Model parameters as in Fig.~\ref{Wells}.}\label{phase_diag}
\end{figure}
Here, $B_x$ controls both the bulk compressibility and the strength of the anisotropy in the periodic phase, while $r$ acts as an effective temperature parameter. Vapor-liquid transformations are controlled by the $\chi^{(3)}$ and $\chi^{(4)}$ functions. These are effective 3- and 4-point correlation functions, given by  $\chi^{(3)}=(a r+b)\chi(\br_1-\br_2)\chi(\br_1-\br_3)$ and $\chi^{(4)}=c\chi(\br_1-\br_2)\chi(\br_1-\br_3)\chi(\br_1-\br_4)$, with $\chi(k)=\text{exp}(-k^2/(2\lambda))$ in reciprocal space. $\chi$ affects low $\textbf{k}$ modes, only picking up density contributions at long wavelengths. The $a$, $b$ and $c$ parameters determine the bulk properties of the uniform phases, while  $\lambda$ affects surface energetics. We present the qualitative physics of the model here, while the study of interface energies will be discussed elsewhere.

Substituting a uniform $n\ofr=n_0$ into eq.~(\ref{free}) yields a Landau free energy in terms of $n_o$ for uniform phases (liquid/vapor). This is shown in Fig.~\ref{Wells}. For simplicity, only 2D results are presented here.
For non-zero $a$, $b$ and $c$ parameters, at low enough rescaled temperature $r$, a double well landscape sets in between liquid and vapor. 
The definition of pressure, $\mathcal{P}=-(\mathcal{F}/V-\mu n_0)$, gives the bulk moduli of the uniform phases  $\beta=n_0\left( \partial \mathcal{P} / \partial n_0\right) $. The vapor and liquid bulk moduli can be made different by several orders of magnitude, consistent with physical systems. For the parameter $r=0.15$,  $\beta_{liq}\sim10^{-3}$ in the liquid region, while in the vapor region, $\beta_{gas}$ varies between $\sim10^{-4}$ in coexistence to  $\sim10^{-6}$ near $n_0\sim0.001$. $\beta$ vanishes as the critical point is approached, where the compressibility diverges with an exponent of $(r-r^*)^{-1}$.
The periodic phase of the functional is treated via a 1-mode approximation~\cite{elder2002}, leading to a Landau theory in both the average density and the amplitude of the solid. Minimizing out the amplitudes gives the solid free energy, a few examples of which are also plotted in Fig.~\ref{Wells}. The phase diagram can be computed by performing common tangent constructions on the Landau theory for different pressures. Fig.~\ref{phase_diag}(a) shows the density-temperature-pressure phase diagram of eq.~(\ref{free}). It features solid-liquid, solid-vapor and vapor-liquid coexistence regions, and is in excellent qualitative agreement with experimental phase diagrams for pure materials~\cite{walas2013,plischkebergersen}. The vapor-liquid phase separation is parabolic, due to the expanded nature of the theory. Higher order long range correlation terms may be added systematically to fine-tune this behaviour. The Pressure-Temperature phase diagram (Fig.~\ref{phase_diag}(b)) also shows a behaviour consistent with experiments. Along with the equilibrium phase boundaries, Fig.~\ref{phase_diag}(b) also shows analytical estimates for the metastability regions of the different phases (dashed lines). Transforming from a metastable to stable phase requires a nucleation event. Crossing the metastable boundaries is associated with the appearance of an unstable wavelength, which spontaneously triggers the phase change, as demonstrated below.

Along with the 1-mode predictions, Fig.~\ref{phase_diag} shows direct simulation results. Simulations involving a periodic phase were initialized as a slab of 1-mode approximation solid in contact with a uniform phase, at the predicted respective average densities. Density was evolved in a $200$ by $2000$ grid point box using eq.~(\ref{denseq}) (discussed below) with a semi-implicit Fourier method, until convergence was reached (See appendix for a explanation of the numerical method). Unless otherwise stated, the grid spacing $dx=a_0/10$ with $a_0$ the lattice constant, time step $dt=1$, $\lambda=0.21$ and $\Gamma=10$ (see figures for other parameters). Fig.~\ref{phase_diag} shows that the coexistence densities (shown in (a)) and pressures (shown in (b)) from direct simulation are in excellent qualitative agreement with our analytical $r$-$n_o$-$P$ and $r$-$P$ space calculations, respectively. Deviations at low average density are in part due to finite size effects, and due to surface energetics not captured in the phase diagram analysis.

\begin{figure}
\centering
\includegraphics[width=0.5\textwidth]{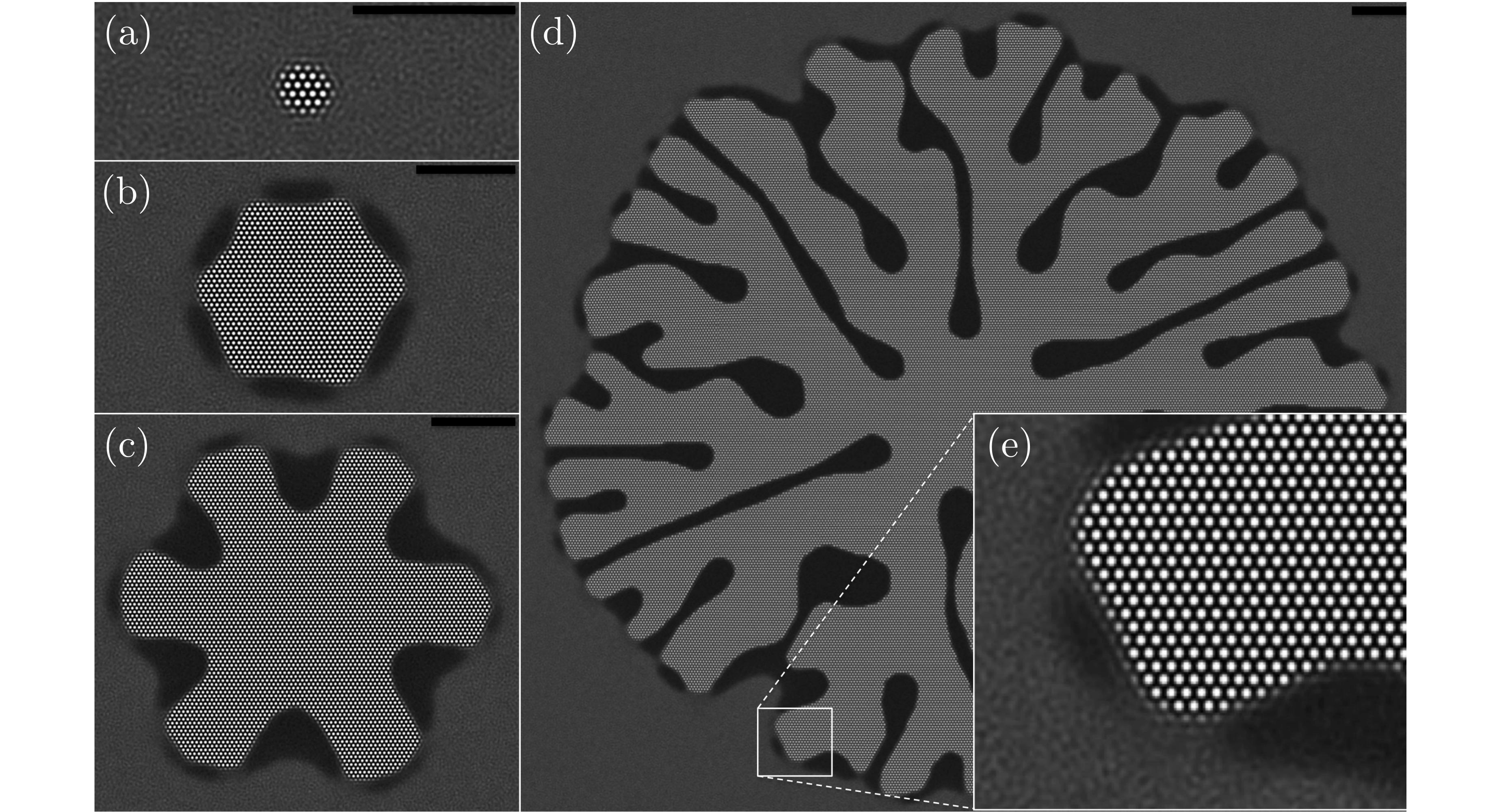}
\caption{3-phase dendritic growth. A solid seed (periodic regions) grows into a metastable liquid (gray uniform areas). High depletion areas nucleate vapor pockets (black regions). a) t=100, b) t=4177, c) t=10293, d) t=36797, e) inset of d). Scale bar: 20 lattice units. Model parameters: $a = 35$, $b = -12.01$, $c= 33.5$, $B_x=0.3$, $n_0=0.125$, $N_a=0.01$, $r=0.145$.\label{dend}}
\end{figure}

To probe the 3-phase kinetics at fixed volume, another simulation was performed where a uniform liquid was quenched into solid-vapor coexistence. The metastable liquid is seeded with a crystal, which grows (Fig.~\ref{dend}(a)). As the solid depletes the surrounding liquid density, vapor pockets nucleate in high depletion areas (Fig.~\ref{dend}(b)). Due to the different growth rates into liquid and vapor, long faceted solid branches are created (Fig.~\ref{dend}(c)), and the resulting structure is a seaweed-like dendrite (Fig.~\ref{dend}(d)).
%

Changes in system volume  $V=dx^2 N_x N_y$ (for a 2D $N_x$ by $N_y$ grid) can be induced by modifying $dx$. As $V$ changes one also modifies the average density, $n_o$,  so that $N=n_0\cdot V$ remains constant. 
In practice this is done by adding a uniform density flux $J_V$ everywhere such as to recover the correct $n_0$.

To control the system pressure, we derived an equation of motion for the volume of the system, that is based on a control algorithm for $\omega'=-\Omega/V$, where $\Omega$ is the grand potential of the system. 
Applying the first law of thermodynamics to an infinitesimally small volume element, enclosed in a larger volume: $ds = (1/T) d e - (\mu/T) d \rho + (P_0/(V T)) d V$, where $T$ is the temperature, $\mu$ the chemical potential, $s$ the entropy density of the volume element, $e$ its internal energy density, $\rho$ the local number density, $V$  the volume of the whole system and $P_0$ is an externally imposed pressure. The natural variables of entropy are $e$, $\rho$ and $V$, and so changes in $\left. \delta s/\delta e \right|_{\mathrlap{\rho,V}}\qquad$, $\left. \delta s/\delta \rho\right|_{\mathrlap{e,V}}\qquad$, $\left. \delta s/\delta V \right|_{\mathrlap{e,\rho}}\qquad$ drive the system. $e$ and $\rho$ obey conservation equations, but assuming an isothermal system, their evolution can be derived from a single density equation, i.e., of the form in eq.~(\ref{denseq}). Volume $V$, considered as a dynamical variable, is a non-conserved global variable and therefore depends on all sub-elements. To linear order in the driving forces,
$
\frac{\partial V}{\partial t} = -M\int_V \text{d}\br  \left( M_{V}\frac{\partial s}{\partial V}+M_{e}\frac{\partial s}{\partial e}+M_{\rho}\frac{\partial s}{\partial \rho}\right)
$
\begin{figure}
\includegraphics[width=0.5\textwidth]{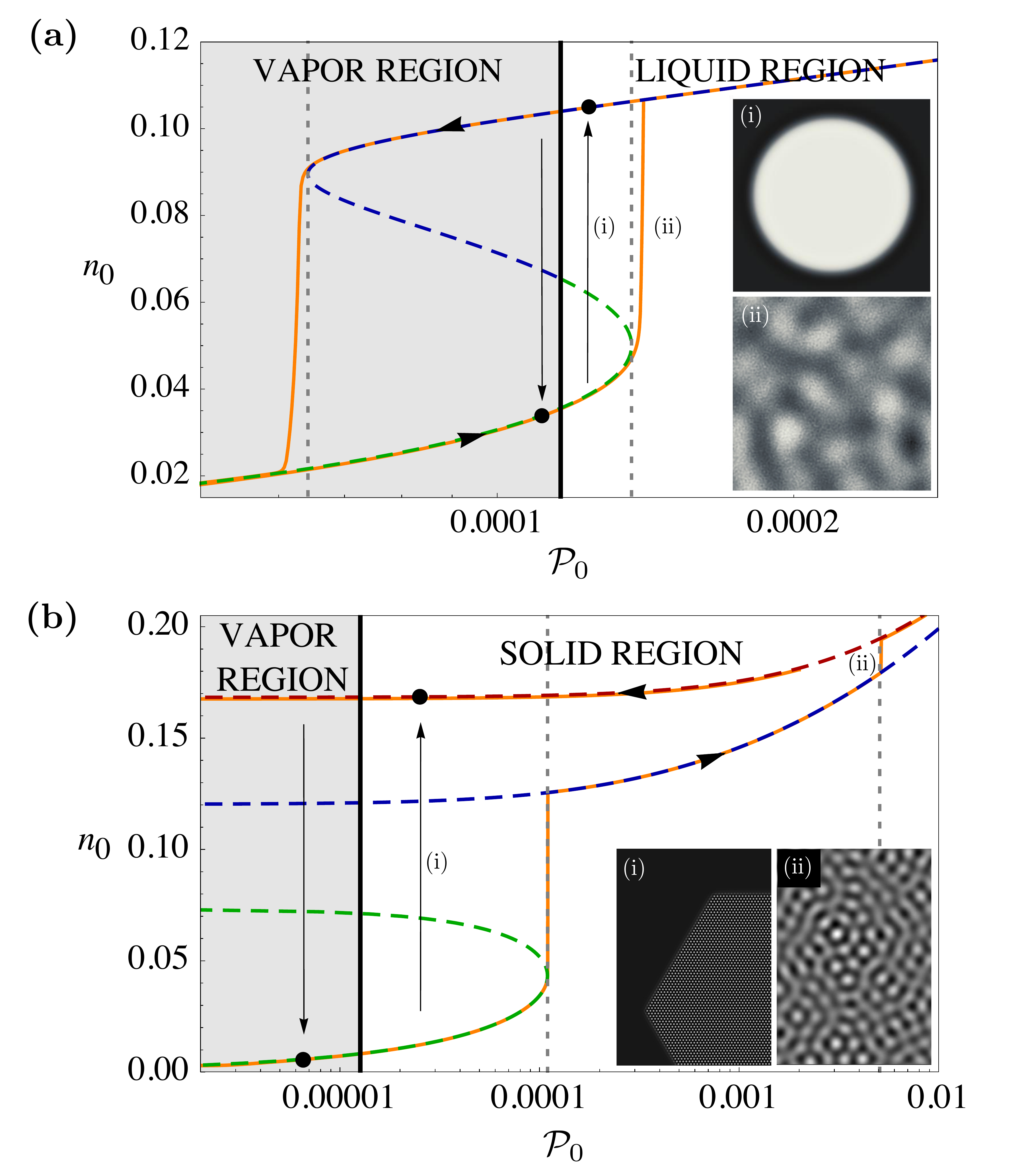}
\caption{Pressure controlled vapor-liquid ((a), $r=0.155$) and vapor-solid ((b), $r=0.147$) transformations. Thick vertical black line: equilibrium condensation/deposition point. Dashed green/blue/red lines: average density vs. pressure for vapor/liquid/periodic phases. Continuous orange lines: system under pressure that is continuously increased/decreased (arrows show direction, $N_a=0.01$). Vertical black arrows: seeded growth of a stable phase out of a metastable phase ($N_a=0$). Insets show snapshots of the order parameter.   Parameters as in Fig.~\ref{dend}.\label{VLVS}}
\end{figure}
where $M_e$, $M_\rho$ and $M_V$ are constants that depend on system variables ($e$, $T$, $s$, $V$, $\rho$...), $1/M$ fixes the timescale of volume changes, while the integral over the system volume ensures a response only to global variations. Using the Gibbs relations, 
$- \left(\partial V /\partial t\right)/M =1/V\int_V\text{d}\br \left\{ M_{V} P_0 +V(M_{e} -M_{\rho} \mu)  \right\}/T$.
Stationarity, $\partial V / \partial t=0$, implies that $V(M_{e} -M_{\rho} \mu) $ should be consistent with a pressure. This condition constrains the expressions for $M_V$, $M_e$ and $M_\rho$. We postulate that $M_V=1/\bar{\rho}k_B$, $M_e= (e-Ts)/(\bar{\rho}k_BV^2)=f/(\bar{\rho}k_BV^2)$, and $M_{\rho}=\rho/(\bar{\rho}k_BV^2)$, so that the final evolution equation reads
\begin{eqnarray}
\frac{\partial n}{\partial t}&=& \Gamma \nabla^2\left(\frac{\delta\mathcal{F}}{\delta n}\right) + N_a\eta \label{denseq} \\
 \frac{\partial V}{\partial t} &=& M (\omega-\mathcal{P}_0) \label{dynP}
\end{eqnarray}
Where $\omega=\int_V \text{d}\br \left(-f+\mu \rho\right)/(\bar{\rho}k_BTV)$ emerges as the adimensional functional generalization of $\omega'$ and $\mathcal{P}_0=P_0/(\bar{\rho}k_BT)$. The noise $\eta$ is a gaussian stochastic variable which satisfies $\langle \eta(\br,t) \eta(\br',t') \rangle =\nabla\cdot\nabla \delta (\br -\br') \delta (t-t')$, with $N_a$ the noise amplitude. Eq.~(\ref{dynP}) is effectively a simple control loop which increases/decreases the volume so that $\omega$ matches the externally imposed pressure $\mathcal{P}_0$, a barostat with timescale $1/M$.

In the absence of defects or interfaces, $\omega$ reduces to the thermodynamic pressure, but in complex bulk solids it additionally convolves interface and strain energies. To demonstrate our formalism, we only consider situations where, transient states aside, $\omega$ tracks pressure. In the particular case of seeded cyrstal growth simulations, interfaces are always present and the $\omega$ integral is therefore restricted to a bulk region where density is uniform. In all constant pressure simulations, $M/(N_xN_y)=2$ (unless otherwise stated), and where the average density increases [decreases], the initial grid spacing was $dx=a_0/8$ $[a_0/35]$.

Simulations of pressure induced transformation were first tested on vapor-liquid systems (Fig.~\ref{VLVS}(a)). The initialization is done in either of the uniform phases, and stabilized to an initial pressure over 5000 time steps, using eqs.~(\ref{denseq}) and (\ref{dynP}) with $\Gamma=10$ on a $1008^2$ grid. The target  pressure $\mathcal{P}_0$ is then ramped up/down continuously, at a rate of $\pm2.7027\cdot10^{-10}$. Because of fluctuations ($N_a=0.01$), the starting phases stay metastable for as long as their compressibility stays positive, before spinodaly decomposing to the equilibrium phase (changes along orange lines in Fig~\ref{VLVS}(a) and inset (ii)). To  illustrate equilibrium transitions, metastable phases are seeded with the equilibrium phase (radius of $300$ grid points), while pressure is monitored in the surrounding bulk. Pressure controlled growth follows (up/down arrows and inset (i)). Once the system is converted, it relaxes at a controlled average pressure.

The vapor-solid transition was tested in a similar manner (Fig.~\ref{VLVS}(b)). Using $M/(N_xN_y)=15$, the vapor phase pressure is continuously increased into the solid region, at a rate of $7.375\cdot10^{-11}$ up to $\mathcal{P}_0=0.00012$, and then a rate of $2.48866\cdot10^{-8}$ to $\mathcal{P}_0=0.01$. As the vapor crosses its metastability region, it spinodally decomposes to a liquid. The liquid then stays metastable until the crystal wavelength becomes unstable, spontaneously triggering another phase change into solid (lower orange line in Fig.~\ref{VLVS}(b), and inset (ii)).  Equilibrium vapor-solid growth is induced by introducing a circular solid seed into the metastable vapor, just past the vapor-solid transition line (upward arrow and inset (i)). The seed first relaxes to a hexagon, and controlling the vapor pressure then leads to a slow layered growth. If the target pressure is below the equilibrium vaporization temperature, the seed sublimates (downward arrow). Due to the absence of unstable boundaries, defects or noise, the bulk solid cannot be vaporized by under pressurizing it (topmost orange line). Well below the equilibrium vaporization point, vapor pockets can remain metastable for a long time due to pinning effects.

The new formalism introduced here allows for novel, and experimentally relevant, applications in solid-vapor growth to be explored. Our theory captures the thermodynamics of pure substances excellently, while maintaining a fundamental connection with all elasto-plastic properties of solids. The formalism introduced here offers new tools to model experimental processes in the fields of crystal growth (chemical vapor deposition or vapor-solid-liquid growth) or soft matter systems (phase separation in polymers, polymer crystals or colloidal suspensions). In this work we demonstrated how to control pressure by changing volume; it is straightforward to control pressure through density changes only, with a suitable replacement for eq.~(\ref{dynP}). While purely technical  issues still remain in regards to controlling pressure directly in complex bulk solids with interfaces and strain, this does not affect the theory. Future work will address the deconvolution of pressure from $\omega$. One approach, for example, is to surround the system with a separate field describing an atmosphere.
\begin{acknowledgments}
The authors thank Nana Ofori-Opoku for useful discussions, The National Science and Engineering Research Council of Canada for funding and Compute Canada for HPC.
\end{acknowledgments}

\section{Appendix: Numerical method}
We start by re-writing the free energy of the system, expanding the 2-point term and re arranging the integrals in the 3- and 4-point terms:
\begin{eqnarray}
\mathcal{F}\left[n\right]&=&\int \text{d}\textbf{r}f'[n]\nonumber\\
&=&\int \text{d}\textbf{r}\left( \frac {n} 2\left[r+B_x(1+\nabla^2)^2\right]n \right.\nonumber\\
&-&\left.\frac {n^3} 6+\frac {n^4} {12}+\frac 1 3 (ar+b)n n_{mf}^2+\frac1 4 c n n_{mf}^3\right)\label{free}
\end{eqnarray}

Where we have introduced $n_{mf}(\br)=\int\text{d}\br'\chi(\br-\br')n(\br)$. This writing of the 3- and 4-point terms explicits the terms as mean field additions, as argued in the paper introduction. In all simulations a semi-implict Fourier space method~\cite{berry2008,provataselder2010}  was used to evolve the order parameter field. For constant volume simulations, only eq.~(2) in the paper needs to be solved. We apply an Euler time stepping scheme in reciprocal space, where the linear terms are implicitly evaluated at $t+\Delta t$ while all the non-linear terms are evaluated at time $t$. Rearranging the terms yields:
\begin{eqnarray}
&n_k(t+\Delta t)&=\\
&&\!\!\!\!\!\!\!\!\!\!\!\!\!\!\!\!\!\!\!\!\!\!\!\!\!\!\!\!\!\frac{ n_k(t)-\Delta t\Gamma \bk^2\left[-\frac 1 2 n^2+ \frac 1 3 {n^3}  + (ar+b)n_{mf}^2+cn_{mf}^3\right]_\bk}
{1+\Delta t\Gamma\bk^2\left(r+B_x\left(1-\bk^2\right)\right)  }\nonumber
\end{eqnarray}
where the $[..]_k$ notation designates the fourier transform of the term in brackets. The volume ($V=dx^2N_xN_y$ for a uniform grid) can be controlled by modifying the grid spacing $dx$, under the constraint that $N= n_0(t)\cdot V(t)$ remains a constant ($n_0(t)$ is the system average of $n(t)$). This is ensured by adding a factor $\left[-n_0(t)+n_0(t_0)dx(t_0)^2/dx(t)^2\right]$ to each grid point, with $t_0$ the initial time.

Running a simulation with barostat control requires eq.~(3) to be solved alongside eq.~(2), this is done by Euler time stepping:
\be
dx(t+\Delta t)=dx(t)+dt\cdot\frac{M}{2dx(t)N_xN_y}(\omega(t)-\mathcal{P}_0)\label{dxchange}
\ee
To avoid over straining the system, the second term in eq.~(\ref{dxchange}) is numerically capped to 0.00003. Depending on the situation, $\omega$ is determined locally or globally. At a point $\br^*$: $\omega_{local}(\br^*,t)=1/V\int\text{d}\br' \left[f'(\br')-n(\br')\delta \mathcal{F}/\delta n(\br')\right]\chi(\br^*-\br')$, here the application of the $\chi(k)=\text{exp}(-k^2/(2\lambda))$ function with $\lambda=0.21$ averages over a region lager than an order parameter periodicity. For a global measurement we use $\omega_{global}(t)=1/V\int\text{d}\br~\omega_{local}(\br) $.


\end{document}